# SoK: Motion Data Privacy in Extended Reality

Azim Ibragimov, Alina Vasina, Uliana Polshcha, Eric D. Ragan
University of Florida
Gainesville, Florida, USA

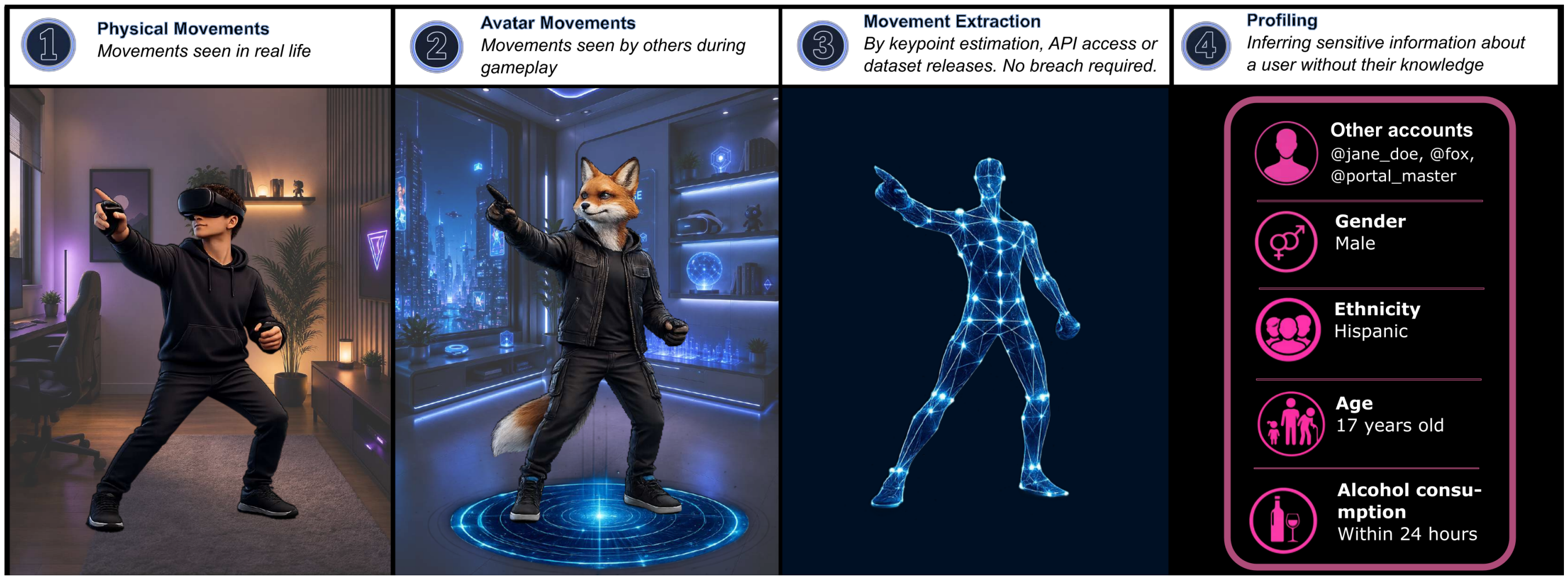

**Figure 1: In Extended Reality (XR), people appear as virtual avatars, which may give users a false sense of anonymity since others cannot see their real faces or bodies. However, XR devices are equipped with motion sensors (i.e., controllers, headset, eye trackers) that reveal the motion pattern of a person behind the virtual avatar. These motion patterns have long been used for surveillance purposes (e.g., gait identification and profiling), and with the widespread adoption of XR systems, they have become easier to obtain than ever. Adversaries can perform privacy attacks by observing a user's virtual avatar, estimating the locations of key joints, accessing recordings in research datasets, or obtaining API access. This SoK examines the privacy concerns posed by these motion patterns and the ease with which they can be collected within XR.**

## Abstract

Extended Reality (XR) provides immersive, interactive 3D experiences. To enable these experiences, the devices must track user motion so the system can respond to actions such as grabbing, looking at, or moving an object. However, motion tracking has raised privacy concerns since it records a person's motion patterns. These motion patterns have been studied extensively across various fields (i.e., gait identification and profiling) and have been shown to reveal sensitive information. With the adoption of XR, these patterns became easier to record and obtain than ever. This creates a fundamental privacy tension: motion tracking enables core XR functionality yet requires users to compromise their privacy. Prior systematization-of-knowledge (SoK) studies on XR privacy have examined the field broadly, with motion-related research distributed across several privacy domains rather than treated as a distinct area of study. However, XR motion privacy has gained significant momentum since the prior SoK, with the literature nearly quadrupling in size and thereby warranting a dedicated systematization of this topic. This SoK examines 134 relevant papers on privacy concerns in motion patterns recorded by XR headsets, including how adversaries can obtain users' motion patterns, the inferences they can draw from them, and methods for protecting users. Based on this review, we synthesize a taxonomy of motion modalities, representations, and inference risks; develop an XR motion threat model; systematize the attack and defense approaches in the XR motion literature; identify gaps in the literature; and provide guidelines for future studies evaluating motion privacy mechanisms. Together, our SoK clarifies the state of XR motion privacy and provides recommendations for future evaluations.

## 1 Introduction

Extended Reality (XR), an umbrella term for Virtual Reality (VR), Augmented Reality (AR), and Mixed Reality (MR), relies on continuous tracking of user movement to create immersive and interactive experiences [24–26, 143]. Tracked motion enables the environment to respond to user actions, such as looking around, navigating virtual environments, manipulating virtual objects, interacting with others, and controlling avatars that represent them. As a result, signals from the eyes, head, hands, body, and face are not secondary sensor streams in XR; they are fundamental to how users perceive, act within, and communicate through these systems. This can create a privacy problem [51, 106, 107, 156] that is easy to underestimate. In XR, users may assume that changing visible identity cues, such as avatar appearance, usernames, or profile information, is enough to remain anonymous. Identification by motion data challenges this assumption. A user may change their avatar, but the avatar still retains the person's movement patterns. These movements may be observed by others during ordinary interactions or accessed by applications through APIs. As a result, even when visible identity cues are concealed, users' motion patterns remain observable and accessible. Research has shown that these patterns can reveal users' identities [51, 107], demographic characteristics [53, 141], health information [118, 166], cognitive [9, 106] and behavioral traits [36, 111]. Thus, a naive user who changes their avatar and username in the hope of remaining anonymous may still be identifiable, and sensitive information about them may be inferred.

Risks posed by motion patterns within XR have been acknowledged by previous surveys and SoKs [2, 11, 15, 19, 35, 37, 44, 46, 112, 157]. Among these, the most relevant SoK by Garrido et al. [35] systematized privacy risks in VR environments across broad privacy domains, including geolocation telemetry, inertial telemetry, text, audio, video, psychological signals, system and network data, and behavior. Motion privacy research appears across several of these categories rather than being treated as a distinct organizing dimension. At the time of their database search (August 2022), this was a natural representation of a field in which motion privacy research was still relatively limited and distributed across different privacy domains. Since then, however, the XR motion privacy literature has nearly quadrupled.

We build on this prior work by examining motion as a distinct organizing dimension of XR privacy. This shift is motivated by both the centrality of movement in XR systems and the rapid growth of the literature. This growth suggests that XR motion privacy is now sufficiently large and technically diverse to warrant dedicated systematization. Our analysis reveals gaps and inconsistencies in evaluation practices that complicate comparison across studies. To address these challenges, we organize the literature using a common threat model and taxonomy and systematically categorize attack and defense research. This paper makes the following contributions:

(1) We provide a systematization of XR motion privacy research covering 134 publications selected, which we synthesize into an XR motion threat model (§4) and taxonomy (§5).
(2) We identify gaps and inconsistencies in the assumptions and evaluation practices used in XR motion privacy research, particularly in utility assessment, adversary modeling, multimodal evaluation, exposure-point analysis, and the treatment of candidate-set size (§6, §7). Based on these findings, we provide recommendations for the design and evaluation of future defenses (§8).
(3) We release the screened-paper list, exclusion decisions, final corpus, and coding sheet, enabling reproduction of our review and supporting future extensions of the XR motion privacy literature.

# 2 Background & Related Work

## 2.1 Extended Reality and Sensors.

Modern wearable XR systems include devices that completely replace one's field of view with a virtual projection (e.g., Meta Quest [92], and HTC Vive Pro 2 [47]), devices that preserve a clear view of the real world while overlaying virtual content (e.g., Google Glass [38] and Ray Ban Meta AI Glasses [94]), and devices that allow users to switch between both modes (e.g., Meta Quest Pro [93] and Apple Vision Pro [3]). These devices project virtual objects into the user's field of view, creating a sense of immersion. XR devices are used for gaming [120], education [4], training [125], and military [70] applications. To enable immersive experiences, natural interaction, presence, XR devices commonly track user motion using sensors built into the devices. These sensors can track hand motion (through controllers or gesture tracking [14]), head motion (through IMU sensors located within the HMD [67]), eye motion (through eye-tracking sensors [11]), full-body motion (through full-body tracking suits [22] or external cameras [18]), and facial expressions (through inward-facing cameras [161] or infrared sensors [20]).

## 2.2 Privacy Risks of Motion Data.

Motion data posed privacy risks long before the widespread adoption of XR [62, 80, 100, 114, 116, 123]. Gait is a prominent example: it can serve as a distinctive behavioral biometric for identity recognition [21, 62, 79, 80, 100, 114, 116, 123] and demographic inference [50, 68, 69, 72, 86, 162]. Historically, however, collecting large amounts of high-quality gait data has been challenging [52] due to factors such as substantial capture distances [133], outdoor environments [144], and varying environmental conditions [140], which can reduce data quality. Recruiting participants for biometric studies can also be difficult [99].

XR substantially reduces these collection barriers. Motion is recorded directly by sensors on or near the user's body, reducing issues related to distance and external recording conditions [14]. Users may also voluntarily upload their motion recordings online [107, 109], while participation in online XR games and social applications can expose movements through avatars even without explicit uploads [91]. Consequently, large-scale XR motion datasets have emerged, including datasets containing recordings from as many as 105,000 users collected from publicly uploaded recordings [109].

The combination of high-quality XR motion sensing and large-scale data availability creates an unprecedented privacy risk by enabling privacy attacks at scale. Prior research has demonstrated the inference of identity, demographic attributes, health information, cognitive characteristics, and behavioral cues from motion data. These inference risks are the primary focus of our systematization.

## 2.3 Relationship to Prior Surveys and SoKs

Recognizing the privacy risks posed by XR headsets, researchers have developed surveys and SoKs that provide an important foundation for the XR privacy field [2, 11, 15, 19, 35, 37, 44, 46, 112, 157]. However, these works either a) examine XR privacy and security broadly [19, 35, 37], b) focus on specific application domains (e.g., healthcare [2], virtual commerce [44]), c) focus on particular motion modalities (e.g., eye tracking [11], body tracking [112]), or d) focus on particular a privacy risk (e.g., recognition [46, 157], health [2]).

For example, the most closely related SoK, by Garrido et al. [35], systematizes the broader VR privacy literature based on 75 publications spanning geospatial telemetry, inertial telemetry, audio, text, video, physiological signals, system and network data, and behavioral information. Motion-relevant research appears across several of these categories rather than being treated as a distinct organizing dimension. At the time of their literature search, conducted in August 2022, this was a natural representation of a field in which motion-related privacy research was still relatively limited and distributed across different privacy domains.

Since then, however, XR motion privacy has developed into a substantially larger and more diverse body of research. As shown in Section 3, the literature meeting our motion-focused inclusion criteria has nearly quadrupled since the search conducted by Garrido et al. [35]. This growth has also introduced new motion modalities, inference targets, threat models, defense mechanisms, and evaluation practices. Consequently, research that previously appeared across broader privacy categories can now be examined as a coherent body of work centered on the privacy implications of motion itself. The growth and maturation of this literature therefore motivate a dedicated systematization of motion data privacy in XR.

## 3 Method: Data Collection & Analysis

Three researchers conducted a structured literature review of technical work on XR motion privacy. Section 3.1 describes the search strategy used to identify the initial corpus. Section 3.2 describes the inclusion criteria and screening procedure used to determine the final corpus. Finally, Section 3.3 describes how the included papers were analyzed and coded to develop the threat model and taxonomy used throughout this work. Throughout this process, the researchers iteratively reviewed papers to discuss interpretations, resolve inclusion and exclusion decisions, and reconcile differences in analysis and coding. Decisions were made through discussion and common agreement whenever possible; in the rare cases where agreement could not be reached (3.73% of decisions), the final decision was determined by majority vote.

### 3.1 Search Strategy

Our hybrid search strategy included database searches (§3.1.1), manual searches (§3.1.2), as well as backward and forward citation searches (§3.1.3).

*3.1.1 Database Search.* For the automatic search, we used IEEE Xplore, the ACM Digital Library, the ACM Guide to Computing Literature, ScienceDirect, and Elsevier on May 15, 2026, with no start-year restriction. These databases cover relevant venues across three areas: (1) XR, with examples including IEEE VR, IEEE ISMAR, IEEE TVCG, ACM VRST, and ACM CHI; (2) biometrics, with examples including IEEE TBIOM, IEEE IJCB, and IEEE FG; and (3) security, with examples including USENIX Security, IEEE S&P, and NDSS. The search query combined three groups of terms: XR terms (*virtual reality*, *augmented reality*, *mixed reality*, *extended reality*, *VR*, *AR*, *MR*, *XR*), motion terms (*motion*, *movement*, *pose*, *tracking*, *telemetry*, *gaze*, *eye*, *head*, *hand*, *controller*, *gesture*, *body*, *skeleton*, *gait*, *locomotion*, *facial*, *avatar*), and privacy or inference terms (*privacy*, *re-identification*, *tracking*, *profiling*, *anonymization*, *obfuscation*, *differential privacy*, *authentication*, *biometrics*, *user modeling*, *behavior analysis*, *attribute inference*, *activity recognition*, *emotion recognition*, *stress detection*, *health inference*). The generic Boolean query was:

```
("virtual reality" OR "augmented reality" OR
"mixed reality" OR "extended reality" OR VR
OR AR OR MR OR XR) AND (motion OR movement OR
pose OR tracking OR telemetry OR gaze OR eye OR
head OR hand OR controller OR gesture OR body
OR skeleton OR gait OR locomotion OR facial
OR avatar) AND (privacy OR re-identification
OR tracking OR profiling OR anonymization OR
obfuscation OR "differential privacy" OR "authentication"
OR "biometrics" OR "user modeling" OR "behavior
analysis" OR "attribute inference" OR "activity
recognition" OR "emotion recognition" OR "stress
detection" OR "health inference")
```

Searches were run over the title, abstract, keyword, or metadata fields supported by each database. The search terms were intentionally broad to account for variation in how XR systems, motion modalities, and privacy-relevant inference tasks are described across research communities. The complete set of records retrieved through these searches is reported in the released materials.

*3.1.2 Manual Search.* The database search covered a substantial number of relevant venues. However, we identified one important venue that was not covered by any of the selected databases: the Privacy Enhancing Technologies Symposium and its proceedings (PETS/PoPETs). This venue publishes a substantial body of research relevant to XR motion privacy and notably, the closest related SoK by Garrido et al. [35] was also published there. We therefore considered it necessary to include this venue in our search strategy. Because PoPETs/PETS does not provide a search interface compatible with our automatic search queries, we conducted a manual search of the venue. As part of this manual search, we reviewed every paper published by the venue since 2000, the first available volume in its online archive [1].

*3.1.3 Backward and Forward Citation Chasing.* To improve the reach of our search, we performed backward and forward citation chasing on all papers that met our inclusion criteria after screening. For each included paper, we reviewed its reference list (backward citation chasing) and the papers that cited it (forward citation chasing) to identify additional relevant studies that were not retrieved through the automatic or manual searches. Newly identified papers were screened using the same inclusion and exclusion criteria.

### 3.2 Screening

After performing the search strategy described in Section 3.1, we identified 9,542 unique records through the automatic search and 1,285 records through the manual search. To ensure that the final corpus remained within the scope of XR motion privacy, we applied the inclusion criteria and screening procedure described below.

*3.2.1 Inclusion Criteria.* We included papers that (1) studied wearable VR, AR, MR, or XR systems with visual displays, or motion representations directly used in such systems, including 3D gaze rays, tracked head or hand poses, skeletal poses, avatar motion, or facial animation parameters; (2) involved motion data, including head, hand, eye, gaze, facial, gesture, body, avatar, or derived behavioral signals; and (3) analyzed a privacy risk, attack, defense, system design, or inference method relevant to motion privacy. These criteria allowed us to include work directly applicable to XR while excluding general sensing research with no clear connection to immersive systems.

[1] https://petsymposium.org/popets/

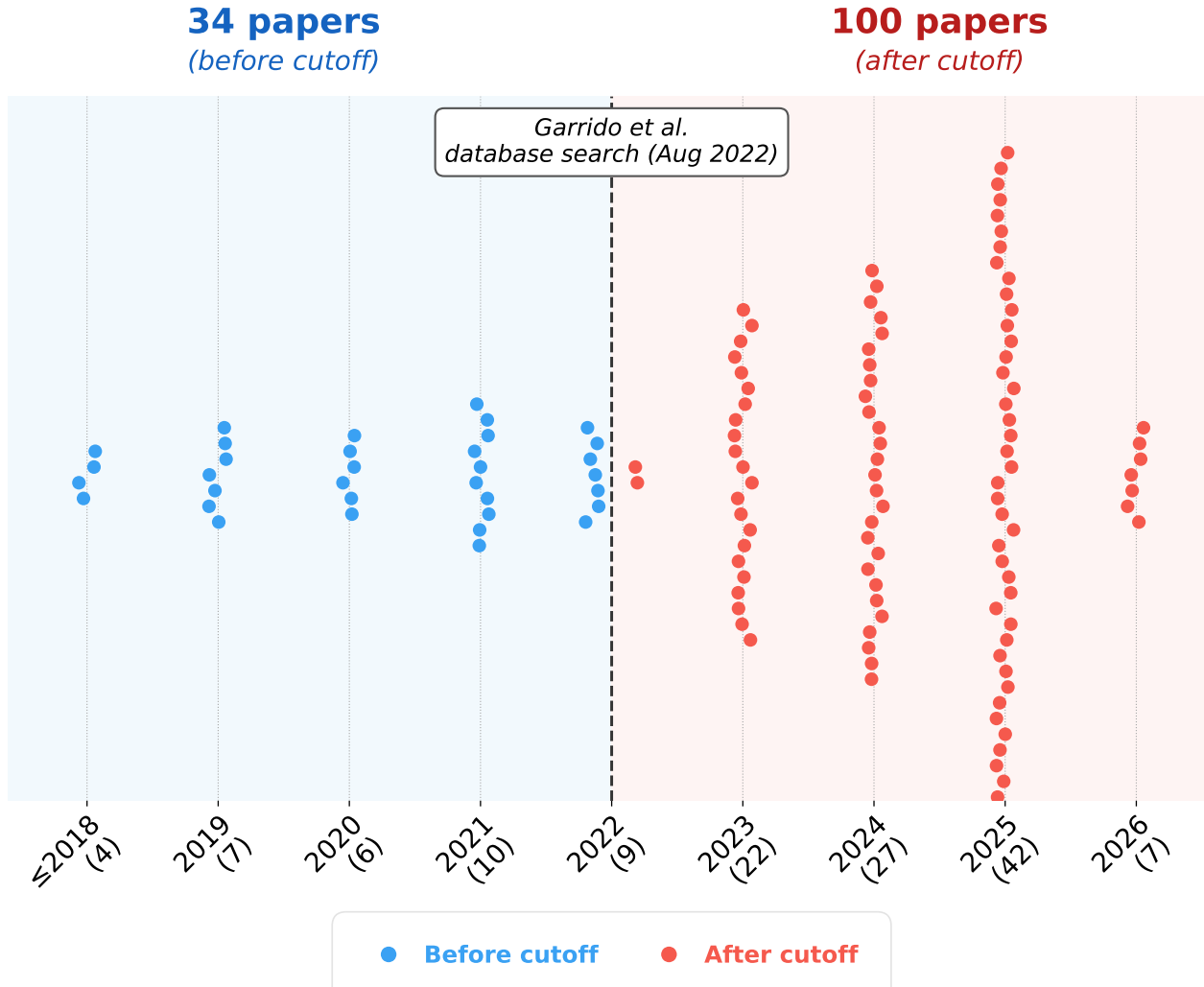

**Figure 2: Publication timeline of the corpus relative to the August 2022 database search of Garrido et al. [35], with annual publication counts shown in parentheses.**

*3.2.2 Screening Procedure.* All records identified through the automatic and manual searches underwent three stages of screening: title screening, abstract screening, and full-text review. At each stage, records that did not satisfy the inclusion criteria described above were excluded. Following title screening, 7,406 records from the automatic search and 48 records from the manual search remained. After abstract screening, 146 and 14 records remained, respectively. Full-text review resulted in the inclusion of 80 papers from the automatic search and 9 papers from the manual search, for a total of 89 papers. We then conducted backward and forward citation chasing on all 89 included papers. Papers identified through citation chasing were evaluated using the same inclusion criteria, resulting in 45 additional papers and bringing the final corpus to 134 papers.

*3.2.3 Timeline of the Corpus.* Figure 2 shows the publication timeline of papers in our corpus, which includes papers published as recently as July 2026, and highlights the substantial growth of XR motion privacy research since the August 2022 database search conducted by Garrido et al. [35]. Applying our inclusion criteria, we identified 34 relevant publications available before this cutoff. Since then, 100 additional relevant papers have been published, bringing the corpus to 134 publications. The literature has nearly quadrupled, suggesting that XR motion privacy has matured into a sufficiently large body of research to warrant dedicated systematization.

## 3.3 Coding and Analysis

After identifying the final corpus, we analyzed and coded the included papers in two stages. In the first stage, we reviewed the corpus to identify recurring concepts, assumptions, and evaluation practices. In the second stage, we used the resulting threat model and taxonomy as a structured codebook for systematically labeling the papers.

*3.3.1 Analysis.* We first reviewed the entire corpus to identify the threat models and assumptions used across the literature. In particular, we analyzed the motion modalities available to adversaries, the exposure points through which adversaries access motion data, the representations and features of motion data, the privacy risks and inference targets considered, dataset sizes, proposed defenses, utility evaluations of these defenses, and the availability of open-science artifacts. After labeling these characteristics for individual papers, we examined the resulting labels across the corpus and iteratively grouped conceptually similar labels into higher-level categories. These groupings were discussed and refined among the researchers until the categories consistently represented recurring concepts observed across the corpus. Based on recurring patterns across the corpus, we developed the XR motion privacy threat model (§4) and the taxonomy (§5).

*3.3.2 Coding.* After completing the initial analysis, we performed a second pass over the corpus to systematically code each paper according to the resulting threat model and taxonomy. We categorized the papers into three groups: *attack*, *defense* and *other*. *Attack* papers propose or evaluate methods that enable privacy attacks (e.g., inferring a user's gender from their motion data). *Defense* papers propose, evaluate, or benchmark privacy-preserving methods by evaluating attacks on protected motion data (e.g., testing whether a user's gender can still be inferred after applying a privacy mechanism). When a paper evaluates attacks on both protected and unprotected data, we classify it as *defense*, as attacks on unprotected data commonly serve as a baseline for evaluating the effectiveness of the defense. The remaining papers, including surveys, SoKs, and dataset papers, were categorized as *other* papers. For both *attack* and *defense* papers, we coded the motion modalities considered and the types of privacy threats addressed. For *defense* papers, we additionally coded the type of privacy mechanism, the number of participants represented in the evaluation, the type of adversary evaluation, the utility evaluation methods, and the availability of open-science artifacts.

# 4 XR Motion Threat Model

Figure 3 presents our XR motion threat model. The model synthesizes assumptions used throughout the literature, including the points at which adversaries can access motion data and the points at which privacy mechanisms can intervene to mitigate privacy risks.

***Adversaries.*** The key adversary assumption of this threat model is that access to motion data does not require a traditional data breach, such as intercepting network traffic or stealing user credentials. Instead, motion becomes available to adversaries as part of ordinary XR sensing, interaction, or data sharing. Based on our analysis of where adversaries obtain motion data across the literature, we identify three adversary types:

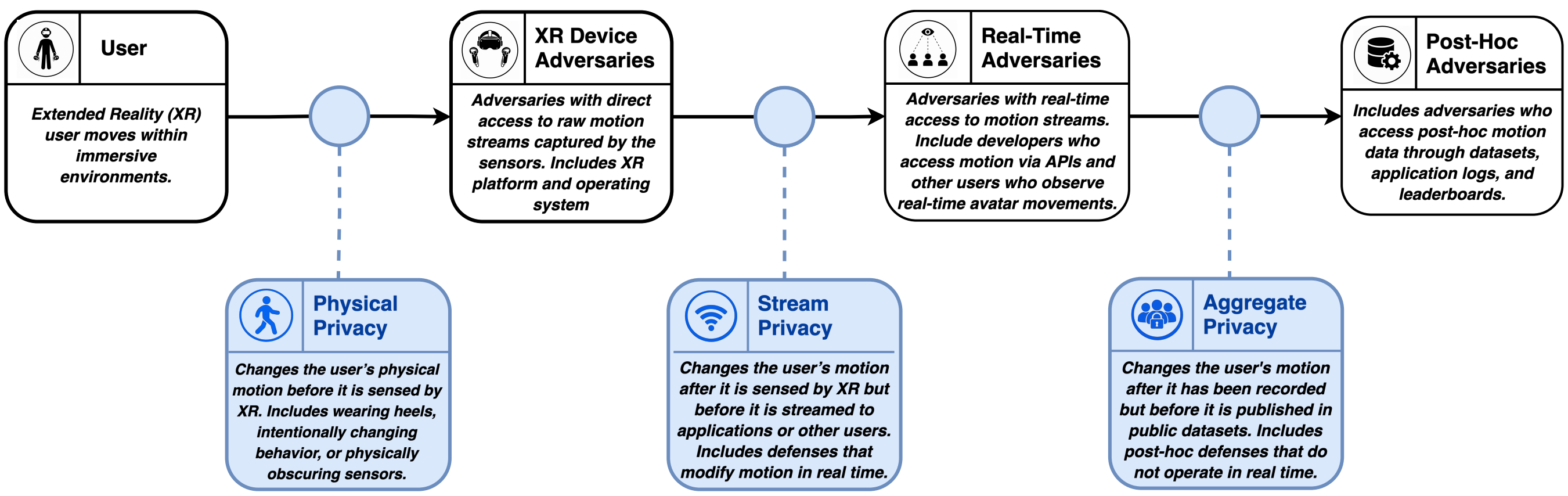

**Figure 3: XR motion privacy threat model. Motion flows from the user through the XR device to real-time and post-hoc adversaries. XR device adversaries have direct access to raw motion streams; real-time adversaries have access to real-time motion streams via APIs or avatar movements; and post-hoc adversaries have access to recorded motion data. The model also identifies three intervention points for privacy mechanisms: physical privacy before sensing, stream privacy before motion is exposed in real time, and aggregate privacy before recorded motion is made available for post-hoc access.**

(A) **XR Device Adversaries.** These adversaries have direct control over XR sensors. This includes the XR device, platform, or operating system. The main strength of this adversary is its privileged position in the motion pipeline: it has access to high-quality motion before any software-based privacy mechanism can be applied and may also control whether such mechanisms are enabled. Its main limitation is accountability. Device and platform providers are typically identifiable entities, and misuse of user data can create legal, regulatory, and reputational consequences.

(B) **Real-Time Adversaries.** These adversaries have access to real-time motion streams during live XR interaction. Developers may obtain motion streams through platform APIs, while other users may record rendered avatar motion and recover movement patterns through keypoint estimation. Their main strength is that access can be difficult to trace: users may not know who recorded their avatar or how application-accessed motion via APIs is later used. Their main limitation is data quality, as developers may receive only selected motion channels and user recordings may be short, occluded, or captured from unfavorable viewpoints.

(C) **Post-Hoc Adversaries.** These adversaries access motion traces after they have already been collected, such as through public datasets, application logs, or leaderboards. Their main strengths are easy access and low traceability: they do not need to collect motion data themselves, and when the data is publicly available, it may be impossible to determine who accessed it and performed the privacy attack. Their main limitation is that they are restricted to the information preserved in the released traces, which may be reduced by missing modalities, lower sampling rates, or preprocessing.

***Privacy Types.*** To protect against the mentioned adversaries, the literature proposes privacy mechanisms that intervene at different points in this motion flow. Based on our analysis of where a mechanism acts relative to sensing, live exposure, and stored motion data, we organize these defenses into three privacy types:

(A) **Physical Privacy** acts before a user's movement is captured by XR sensors and is best suited for users who do not trust their XR device. These mechanisms alter the user's physical movement or interfere with how it is sensed, for example by changing gait through footwear, obscuring eye-tracking sensors, or intentionally modifying behavior. Because protection is applied before sensing, it can protect against all three adversary types. Its main strength is that no downstream party receives the original motion. Its main weakness is the burden on the user, including discomfort, fatigue, behavioral effort, reduced tracking precision, or degraded interaction quality.

(B) **Stream Privacy** acts after motion has been captured but before it is exposed during live interaction and is best suited for users who trust their XR device but not applications or other users. These mechanisms modify the motion stream before it is exposed through application APIs or avatar behavior. Its main strength is that it protects users during ordinary XR interaction without requiring them to physically alter their movement. Its main weakness is that it cannot protect against XR device adversaries that access the original motion before privatization. It also faces strict real-time constraints, as protection must preserve responsiveness, tracking quality, and user comfort.

(C) **Aggregate Privacy** acts after motion data has been collected and is best suited for protecting stored or released motion traces, such as datasets, logs, recordings, or leaderboards. Its main strength is that it operates post hoc and therefore avoids real-time constraints, allowing more computationally expensive transformations or formal privacy guarantees such as differential privacy. Its main weakness is that protection begins only after collection, so XR device and real-time adversaries that already accessed the original motion remain outside its protection.

# 5 Taxonomy of XR Motion Data Privacy Risks

Figure 4 presents our taxonomy of XR motion data and its associated privacy risks. We organize the taxonomy around three components: motion modalities, their representations, and the privacy risks associated with them. Based on our analysis (§3.3), we group the motion modalities observed across the literature into five modalities:

- **Head Motions.** Changes in the position or orientation of the user's head [87, 105, 135], typically captured through the head-mounted display (HMD) [89]. Head motion may be represented as position [6, 51, 107, 135], orientation [51, 107, 135], velocity [87, 105], or acceleration [105].
- **Eye Motions.** Motion and gaze behavior produced by the user's eyes [83–85], including gaze position [142], orientation [10], fixations [29], saccades [29], pupil diameter [158], blink rate [10], scanpaths [119], velocity [83], and acceleration [27].
- **Facial Motions.** Movements of facial features from expressions and other changes in facial configuration [54, 82, 155], represented through facial keypoints [155] or blendshapes [54, 82].
- **Hand Motions.** Motion of the user's hands during XR interaction [14, 74, 111], captured through tracked controllers as position [107], orientation [108], velocity [122], or acceleration [97], or directly through hand tracking as hand keypoints [74].
- **Full-Body Motions.** Motion of the user's body beyond separately tracked head, eye, facial, or hand movements, represented through skeletal joints spanning the body, including avatar joint positions [33], orientations [41], velocities [17], or corresponding joint measurements from full-body tracking suits [90].

For each modality, the taxonomy shows how the motion is represented in XR systems and which representations have been used in the literature for privacy attacks. Finally, we identify the privacy risks associated with each modality. Across the literature, we identify five distinct categories of privacy risk:

- **Recognition.** Includes recognizing, linking, verifying, or re-identifying users from XR motion data. For example, an adversary may determine that multiple seemingly anonymous accounts belong to the same person based on their motion patterns [51, 52, 107, 156].
- **Demographics.** Includes inference of demographic characteristics such as gender, age, race, income, education, and related attributes [13, 53, 134].
- **Health & Physiology.** Includes inference of health conditions, impairments, and physiological information such as heart rate, blood pressure, respiration rate, and other vital signs [88, 118, 166].
- **Cognitive & Affective States.** Includes inference of users' cognitive or affective states, such as attention, stress, emotion, or engagement [9, 44].
- **Behavior, Intent & Preference.** Includes inference of users' behavior, intentions, and preferences, such as future actions, purchase intent, product interest, or other user preferences [36, 111, 121].

Overall, the taxonomy shows that XR privacy risks span multiple motion modalities and representations. Head, eye, and hand movements cover all five privacy-risk categories, while facial and full-body movements have been studied across fewer risks. Together with the threat model (§4), we use this taxonomy to code the corpus in §6 and §7 by motion modality, privacy risk, adversary, and privacy type, and to identify research gaps through the corresponding research questions.

# 6 XR Attacks

Table 1 summarizes the 62 attack papers in our corpus that met the screening criteria (§3.2). For each paper, the table reports the studied motion modality and privacy risk according to our taxonomy (§5). Additionally, the research questions below are designed to address common questions about the attack literature and identify gaps in existing work.

**(RQ1)** ***Which motion modalities do XR motion attacks most often exploit?*** Across the 62 attack papers, head movement is the most frequently exploited modality (40 papers), followed by hand movement (28 papers) and eye movement (28 papers). Full-body motion (3 papers) and facial movement (1 paper) are considerably less represented. This pattern likely reflects the history and availability of XR sensing: head and hand tracking have been standard on commodity VR devices for much longer, whereas full-body and facial tracking became available later or often require more specialized hardware [61]. The attack literature is therefore concentrated around motion streams that are more commonly exposed by XR devices, while newer or less widely available modalities remain comparatively underexplored.

**(RQ2) What privacy risks are the most studied?** Across the 62 attack papers, recognition is the most studied privacy risk (49 papers), followed by health & physiology (8 papers), demographics (8 papers), cognitive & affective states (7 papers), and behavior, intent & preference (3 papers). Recognition therefore dominates the literature. One reason may be its direct security implications. Motion-based recognition can increase the impact of an account compromise beyond a single account [5]. For example, if the password of one XR account is leaked, an attacker could use motion associated with that account to identify other accounts belonging to the same user. If the user reused the same password across accounts, the attacker could then try the leaked password on the newly identified accounts and potentially compromise them as well. While recognition has direct security implications, other privacy risks can also be highly invasive, with consequences such as user profiling, monitoring, and discrimination [106]. Their comparatively limited representation in the literature therefore reveals an important research gap.

**(RQ3) Can an adversary infer multiple privacy risks from the same motion data?** Yes. The same motion data can support multiple privacy inferences, and prior work shows that similar attack pipelines can often be repurposed for different inference targets. In many supervised-learning settings, the motion representation and learning algorithm can remain unchanged while only the prediction labels are changed. For example, a recognition model trained to predict participant identity can instead be trained to predict a demographic attribute such as gender by replacing participant-ID labels with gender labels [106, 111, 141]. Across our attack corpus, 7 papers consider at least two privacy risks. The largest number of risks considered by a single study is four. Nair et al. [111] evaluate all privacy-risk categories except cognitive & affective states, while another study by Nair et al. [106] evaluates all categories except behavior, intent & preference. Thus, no work in our corpus evaluates all five privacy risks within the same study. This represents an important research gap. A privacy mechanism that protects against recognition may still leave users vulnerable to demographic, health, cognitive, or behavioral inference. Evaluating attacks across multiple privacy risks is therefore important for understanding whether a defense provides comprehensive protection rather than protection against only a single inference target.

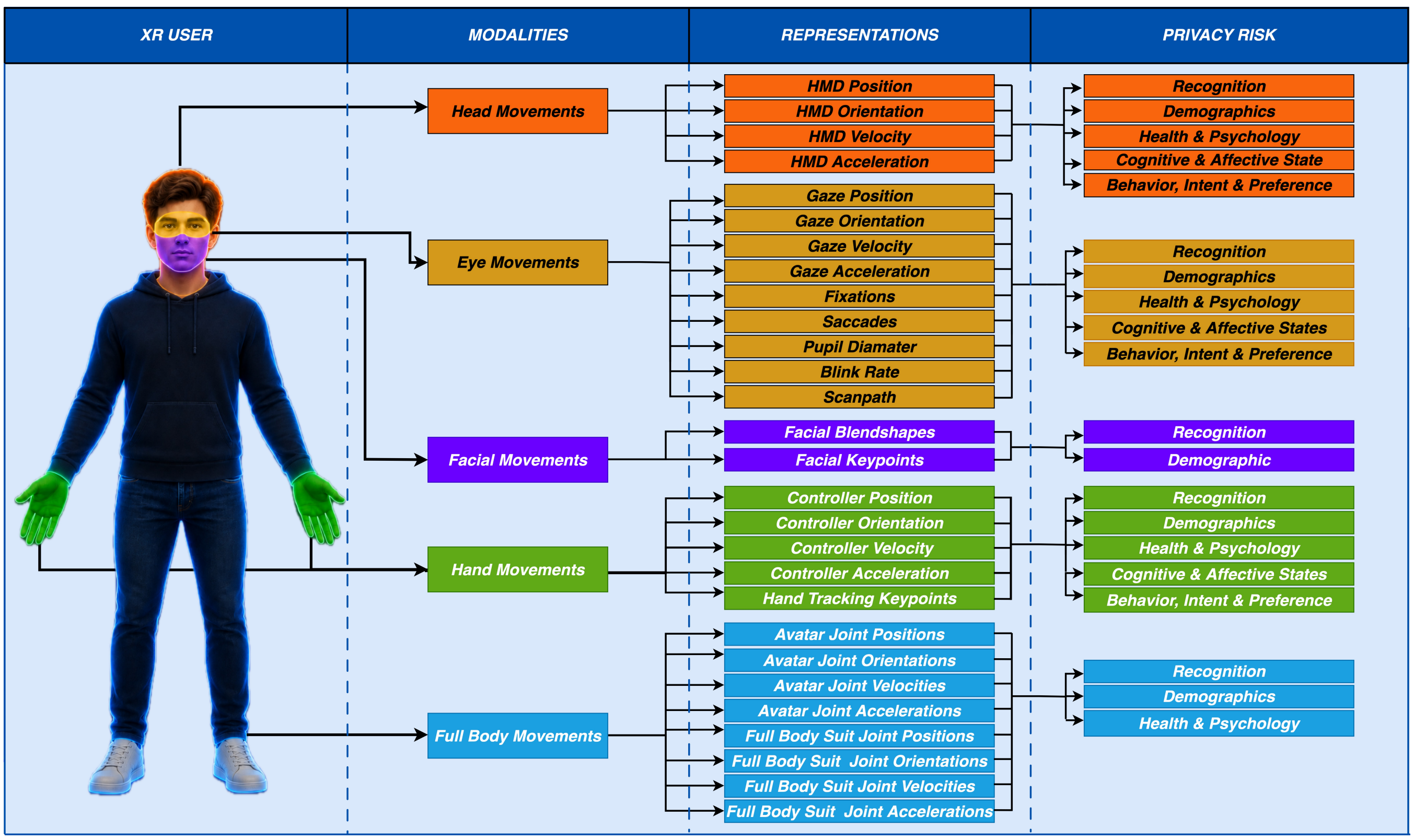

**Figure 4: Taxonomy of XR motion data and inference risks derived from our analysis of the literature. The figure maps XR motion modalities to their representations and five privacy risk categories: recognition; demographics; health and physiology; cognitive and affective states; and behavior, intent, and preferences.**

**(RQ4) Can an adversary have access to multiple motion streams at the same time?** Yes, and this is a common setup in the literature. Across the 62 attack papers, 30 evaluate at least two motion modalities. The most common exact combination is head and hand motion (19 papers). The largest number of modalities considered together by a single study is four, with Jarin et al. [54] evaluating eye, head, hand, and facial movements. Access to multiple modalities can also make an adversary more capable. Prior work shows that multimodal attacks can outperform their single-modality counterparts [6, 51, 141], allowing adversaries to achieve higher attack accuracy. Another concern is avoidance: if a privacy mechanism protects only one modality while others remain exposed, an adversary can simply use an unprotected modality instead [6, 51]. For example, if eye motion is protected but head motion remains unprotected, the adversary can ignore eye motion and perform the attack using head motion. Despite these risks, no work in our corpus considers all five motion modalities simultaneously. This represents an important gap for both attack and defense research.

## 7 XR Defense

Table 2 summarizes the 50 papers in our corpus that that met the screening criteria (§3.2). The defense papers are coded by protected modality, privacy risk they protect against, privacy type, adversary evaluation, utility evaluation, and open-science support. The research questions below are designed to address common questions about the defense literature and identify gaps in existing work.

**Table 1: Systematization of 62 *attack* papers across the XR motion privacy landscape. Each row represents one paper and reports the motion modalities (•) used in the studied privacy attack and the privacy risks (•) demonstrated.**

| | | Attack Modality | | | | | Privacy Risk | | | | |
|---|---|---|---|---|---|---|---|---|---|---|---|
| ID | Attack Paper | Eye Motions | Head Motions | Hand Motions | Full Body Motions | Facial Motions | Recognition | Demographics | Health & Phys. | Cogn. & Affect. | Behav., Intent & Pref. |
| A1 | Zhang et al. [168] | • | | | | | • | | | | |
| A2 | Mustafa et al. [105] | | • | | | | • | | | | |
| A3 | Pfeuffer et al. [122] | • | • | • | | | • | | | | |
| A4 | Bozkir et al. [9] | • | | | | | | | | • | |
| A5 | Liebers et al. [78] | • | | | | | • | | | | |
| A6 | Yan et al. [158] | • | | | | | • | | | | |
| A7 | Sivasamy et al. [135] | | • | | | | • | | | | |
| A8 | Bhalla et al. [8] | | • | • | | | • | | | | |
| A9 | Falk et al. [33] | | • | • | • | | • | | | | |
| A10 | Liebers et al. [73] | | • | • | | | • | | | | |
| A11 | Liebers et al. [76] | • | • | | | | • | | | | |
| A12 | Wierzbowski et al. [154] | • | • | | | | • | | | | |
| A13 | Zhang et al. [166] | | • | • | | | • | | • | | |
| A14 | Miller et al. [98] | | • | • | | | • | | | | |
| A15 | LaRubbio et al. [66] | • | | | | | • | | | | |
| A16 | Jang et al. [53] | | • | • | • | | • | • | • | | |
| A17 | Aziz et al. [5] | • | | | | | • | | | | |
| A18 | Mai et al. [87] | | • | | | | • | | | | |
| A19 | Meng et al. [91] | | • | • | • | | • | | | | |
| A20 | Lohr et al. [85] | • | | | | | • | | | | |
| A21 | Liebers et al. [75] | | • | • | | | • | | | | |
| A22 | Zhang et al. [167] | | • | | | | • | • | • | | |
| A23 | Suzuki et al. [139] | | | • | | | • | | | | |
| A24 | Tricomi et al. [141] | • | • | • | | | • | • | | | |
| A25 | Lohr et al. [84] | • | | | | | • | | | | |
| A26 | Nair et al. [111] | | • | • | | | • | • | • | | • |
| A27 | Liebers et al. [77] | | • | • | | | • | | | | |
| A28 | Jiao et al. [57] | • | | | | | • | | | | |
| A29 | Dossett et al. [31] | | • | | | | • | | | | |
| A30 | Zhao et al. [169] | • | • | | | | • | | | | |
| A31 | Raju et al. [127] | • | | | | | • | | | | |
| A32 | He et al. [45] | | • | | | | • | | | | |
| A33 | Zafar et al. [163] | | • | • | | | • | | | | |
| A34 | Premkumar et al. [124] | | • | • | | | • | | | | |
| A35 | Peng et al. [121] | • | | | | | | | | | • |
| A36 | Jeon et al. [56] | | • | • | | | • | | | | |
| A37 | Miguel-Alonso et al. [95] | • | • | • | | | • | | | | |
| A38 | Rubo et al. [132] | • | | | | | • | | | | |
| A39 | Shadin et al. [134] | | • | • | | | | • | | | |
| A40 | Liebers et al. [74] | | • | • | | | • | | | | |
| A41 | Zhang et al. [165] | | • | | | | • | | • | | |
| A42 | Gyreyiri et al. [40] | | • | | | | • | | | | |
| A43 | Jeon et al. [55] | | • | • | | | • | | | | |
| A44 | Rogers et al. [130] | • | • | | | | • | | | | |
| A45 | Ye et al. [159] | | • | | | | | | • | | |
| A46 | Genco et al. [36] | • | | | | | | | | | • |
| A47 | Wang et al. [146] | | • | • | | | | • | | | |
| A48 | Vaitheeshwari et al. [142] | • | | | | | | | • | | |
| A49 | Wang et al. [145] | | • | • | | | | • | | | |
| A50 | Moore et al. [103] | | • | • | | | • | | | | |
| A51 | Orlosky et al. [118] | • | | | | | | | • | | |
| A52 | Munoz et al. [104] | | • | | | | | | • | | |
| A53 | Zhang et al. [164] | • | | | | | | | • | | |
| A54 | Jarin et al. [54] | • | • | • | | • | • | | | | |
| A55 | Nair et al. [106] | | • | • | | | • | • | • | • | |
| A56 | Normoyle et al. [117] | | • | • | | | • | | | | |
| A57 | Nair et al. [107] | | • | • | | | • | | | | |
| A58 | Raju et al. [126] | • | | | | | • | | | | |
| A59 | Yu et al. [160] | | | • | | | • | | | | |
| A60 | Cheng et al. [23] | • | | | | | | | | • | |
| A61 | Hasan et al. [42] | • | | | | | | | | • | |
| A62 | Lohr et al. [83] | • | | | | | • | | | | |

**(RQ5) What are the most protected modalities and privacy risks?** Across the 50 defense papers, eye motion is the most frequently protected modality (28 papers), followed by head motion (27 papers) and hand motion (21 papers). Full-body motion (7 papers) and facial motion (2 papers) receive substantially less attention. This distribution closely follows the attack literature, where eye, head, and hand motion are also the most frequently studied modalities. A similar pattern appears across privacy risks. Recognition is protected most frequently (42 papers), followed by behavior, intent & preference (8 papers), health & physiology (6 papers), demographics (6 papers), and cognitive & affective states (1 paper). This mirrors the attack literature, where recognition is the most frequently studied privacy risk. Overall, the defense literature reflects many of the same gaps observed on the attack side. Full-body and facial motion remain comparatively underexplored, while defenses are heavily concentrated on recognition rather than the broader range of privacy risks that XR motion can reveal. This leaves limited evidence that existing mechanisms can protect less-studied motion modalities or provide protection against privacy risks beyond recognition.

**(RQ6) *Which privacy types receive the most research attention?*** Across the 50 defense papers in our corpus, research attention is unevenly distributed across the three privacy types defined in our threat model (§4). Stream privacy is the most commonly studied (27 papers), followed by aggregate privacy (20 papers). In contrast, physical privacy is studied far less frequently (3 papers). This distribution reveals a substantial gap in the current defense literature. Physical privacy is the only privacy type in our threat model that can protect users against XR device adversaries, yet it is the least studied. Existing work therefore concentrates primarily on protecting motion after it has already been captured by the XR device, while comparatively little research considers mechanisms that protect users before sensing occurs. This suggests that defenses against adversaries with direct access to XR sensing remain an important direction for future work.

**(RQ7) How does candidate population size impact the comparison and reproduction of defense evaluations?** Candidate population size varies substantially across the defense literature, ranging from 4 to 56,082 participants, with a median of 61 and a mean of 1534. This variation complicates comparison because identification accuracy inherently depends on the number of candidate identities [52, 96]. For example, 50% accuracy among two candidates is equivalent to random guessing, whereas 50% among 55,000 candidates is dramatically above the chance baseline of approximately 0.0018%. Thus, similar attacks or defenses may report substantially different accuracies simply because they are evaluated on different candidate population sizes. This also creates a reproducibility challenge when large datasets are not publicly available. Even when the data-collection and evaluation procedures are fully described, other researchers may be unable to collect a comparably large population. They may reproduce the mechanism itself, but evaluating it on a smaller population can produce different identification results. Rather than requiring a particular population size, we recommend reporting privacy performance across multiple candidate-set sizes. Lohr et al. [84], for example, use bootstrapping to estimate biometric performance as the number of candidate identities varies. A dataset of 300 participants could similarly be evaluated at $N = 50$, $N = 100$, $N = 150$, and so forth, providing common evaluation points across datasets of different sizes. However, Lohr et al. [84] are the only work in our corpus to explicitly evaluate performance as a function of candidate population size, revealing an important methodological gap in current XR motion privacy evaluations.

**Table 2: Systematization of 50 XR motion privacy *defense* papers. Each row reports the motion modalities (•) and privacy risks (•) protected by the proposed defense, together with its privacy type (•), participant population, adversary evaluation (•), utility evaluation (•), and public artifact availability (•).**

| | | Defense Modality | | | | | Privacy Risk | | | | | Privacy Tier | | | Study Scale | Adversary Evaluation | | | Utility Evaluation | | | | Open Science |
|---|---|---|---|---|---|---|---|---|---|---|---|---|---|---|---|---|---|---|---|---|---|---|---|
| ID | Defense Paper | Eye Motions | Head Motions | Hand Motions | Full Body Motions | Facial Motions | Recognition | Demographics | Health & Phys. | Cogn. & Affect. | Behav., Intent & Pref. | Physical | Stream | Aggregate | Participants | Oblivious Adv. | Adaptive Adv. | Models Tested | Task Utility | Spatial Prec. | Temporal Prec. | User Study | Artifact |
| D1 | Steil et al. [136] | • | | | | | • | • | | | | | | • | 20 | | • | 1 | • | | | | |
| D2 | David-John et al. [27] | • | | | | | • | | | | | | • | | 247 | | • | 1 | • | • | | | • |
| D3 | Gordon et al. [39] | • | • | • | | | | | | | • | • | | | 55 | | • | 2 | | | | • | • |
| D4 | Moore et al. [103] | | • | • | | | • | | | | | | | • | 60 | | • | 3 | | | | | • |
| D5 | David-John et al. [29] | • | | | | | • | | | | | | | • | 298 | • | | 1 | • | | | | • |
| D6 | Hu et al. [48] | • | | | | | | | | | • | | • | | 50 | | | 0 | • | • | • | | |
| D7 | Moore et al. [102] | | • | • | | | • | | | | | | | • | 45 | • | | 3 | | | | | • |
| D8 | David-John et al. [28] | • | | | | | • | | | | | | | • | 73 | • | | 1 | • | | | | |
| D9 | Meng et al. [90] | | • | • | • | | • | | | | | | • | | 10 | • | | 3 | | • | | | |
| D10 | Nair et al. [108] | | • | • | | | • | | | | | | • | | 1000 | • | • | 3 | • | | | • | • |
| D11 | Nair et al. [110] | | • | • | | | • | | | | | | • | | 500 | • | | 1 | | | | | |
| D12 | Miller et al. [97] | | • | • | | | • | | | | | | • | | 183 | • | | 1 | | | | | |
| D13 | Wei et al. [152] | | • | • | | | • | | | | | | • | | 24 | • | | 1 | | • | | • | |
| D14 | Sun et al. [138] | | • | • | • | | • | | | | | | | • | 4 | • | | 2 | | • | | | |
| D15 | Wilson et al. [156] | • | | | | | • | | | | | | • | | 44 | • | • | 1 | • | | | • | • |
| D16 | Warin et al. [148] | | • | • | | | • | | | | | | • | | 0 | | | 0 | | | | | |
| D17 | Aziz et al. [6] | • | • | • | | | • | | | | | | • | | 38 | • | • | 1 | | | | | |
| D18 | Guthula et al. [115] | | • | • | | | • | | | | | | | • | 2598 | • | | 1 | • | | | | |
| D19 | Sun et al. [137] | | • | • | | | • | | | | | | | • | 2049 | • | | 2 | | • | | | |
| D20 | Kundu et al. [64] | • | • | | | | • | | • | | | | | • | 34 | | • | 5 | • | | | • | |
| D21 | Ahmed et al. [1] | • | | | | | | • | | | | | | • | 36 | • | | 1 | • | | | | |
| D22 | Ibragimov et al. [51] | • | • | • | | | • | | | | | | • | | 445 | • | • | 1 | • | • | | | • |
| D23 | Ren et al. [129] | • | | | | | • | | | | | | | • | 432 | • | | 1 | • | • | | | |
| D24 | Nair et al. [113] | | • | • | | | • | • | • | | | | • | | 56,082 | • | | 3 | | | | | • |
| D25 | Elfares et al. [32] | • | | | | | • | | | | | | • | | 100 | • | | 1 | • | | | • | • |
| D26 | Li et al. [71] | • | | | | | • | | • | | | | • | | 11 | • | | 1 | • | • | • | • | |
| D27 | Liu et al. [82] | | | | | • | • | | | | | | • | | 45 | | • | 8 | | • | • | • | • |
| D28 | Kundu et al. [63] | • | • | | | | • | | | | | | • | | 94 | • | • | 2 | • | | • | • | |
| D29 | Hasan et al. [43] | • | | | | | | | • | • | | | • | | 322 | | | 0 | • | • | | | |
| D30 | Raju et al. [128] | • | | | | | • | | | | | | • | | 322 | • | | 1 | • | • | • | | |
| D31 | Aziz et al. [7] | • | | | | | • | | | | | | • | | 322 | • | | 1 | • | • | | | |
| D32 | Wilson et al. [155] | | | | | • | • | | | | | | • | | 7005 | • | | 1 | • | • | | | • |
| D33 | David-John et al. [60] | • | | | | | • | | | | | | • | | 15 | • | | 2 | • | • | | | |
| D34 | Liu et al. [81] | • | | | | | • | | • | • | • | | | • | 5 | | | 0 | • | • | | | |
| D35 | Bozkir et al. [10] | • | | | | | • | • | | | | | | • | 37 | • | | 4 | • | • | | | |
| D36 | Wei et al. [153] | • | • | • | | | • | | • | | | | • | | 24 | | • | 2 | | | | | • |
| D37 | Wei et al. [151] | | • | | | | • | | | | • | | • | | 30 | | | 0 | • | | • | | • |
| D38 | Wei et al. [150] | | • | | | | • | | | | • | | • | | 62 | | • | 3 | • | • | | | |
| D39 | Wei et al. [149] | | • | | | | | | | | • | | • | | 48 | | | 0 | • | • | | | |
| D40 | Wang et al. [147] | • | | | | | | | | | • | | • | | 20 | • | | 1 | • | • | • | | |
| D41 | Ozdel et al. [119] | • | | | | | | | | | • | | | • | 91 | | | 0 | • | | | | • |
| D42 | Romero et al. [131] | | • | • | • | | • | | | | | | • | | 20 | • | | 2 | • | • | | • | |
| D43 | Bozkir et al. [12] | • | | | | | | | | | • | | • | | 0 | | | 0 | • | • | | | |
| D44 | John et al. [59] | • | | | | | • | | | | | • | | | 5 | • | | 1 | • | • | | | |
| D45 | John et al. [58] | • | | | | | • | | | | | • | | | 54 | • | | 1 | • | • | | • | |
| D46 | Carr et al. [17] | | • | • | • | | • | | | | | | | • | 146 | • | | 2 | • | | | | • |
| D47 | Carr et al. [16] | | • | • | • | | • | • | • | | | | | • | 146 | • | | 2 | • | • | | | • |
| D48 | Moon et al. [101] | | • | • | • | | • | • | | | | | | • | 140 | | • | 3 | • | | | | • |
| D49 | Hanisch et al. [41] | | • | • | • | | • | | | | | | | • | 162 | | • | 2 | • | | | • | • |
| D50 | Kundu et al. [65] | • | • | | | | • | | | | | | | • | 94 | | • | 7 | • | | | | |

**(RQ8) How do adversary assumptions impact the comparison and reproduction of defense evaluations?** Adversary assumptions significantly impact reported defense performance. In our coding, we distinguish between two evaluation settings: oblivious and adaptive adversaries. An oblivious adversary is unaware that a modality has been privatized and has no knowledge of how the privacy mechanism operates. In contrast, an adaptive adversary knows which modality is protected and understands how the mechanism works. Such an adversary may train attack models directly on privatized data, attempt to denoise protected motion, or ignore protected modalities and exploit unprotected ones instead [51, 108, 156]. Prior work reports substantially different performance under adaptive and oblivious adversaries, with adaptive adversaries consistently achieving higher attack success [108, 155]. Across the defense corpus, 26 papers evaluate only oblivious adversaries, 11 evaluate only adaptive adversaries, and only 5 report results for both adversary types. This creates an important challenge for comparing defense evaluations. A defense evaluated only against an oblivious adversary cannot be directly compared with one evaluated only against an adaptive adversary, since the reported privacy performance corresponds to different threat assumptions. Likewise, reproducing a published evaluation requires reproducing not only the privacy mechanism but also the adversary model under which it was evaluated. The limited use of paired oblivious and adaptive evaluations therefore represents an important gap in the XR motion privacy literature. Since the two adversaries can yield substantially different privacy outcomes, the absence of either evaluation setting makes it difficult to compare reported privacy performance across studies that adopt different adversary assumptions.

**(RQ9) Do utility evaluations match the defense's deployment setting?** We distinguish four types of utility evaluation: task utility (whether privatized motion remains useful for a specific task, such as preserving game scores or area-of-interest accuracy), spatial utility (how far privatized motion deviates from the original motion), temporal utility (processing delay or latency), and user studies (how users experience the mechanism during actual interaction). Task, spatial, and temporal utility can all be measured without requiring users to experience the defense (e.g., by applying the mechanism to recorded motion and computing the corresponding metrics afterward). In contrast, user studies require participants to interact with the defense directly. The appropriate utility evaluation depends on the defense's deployment setting. For aggregate-level privacy, task, spatial, and temporal utility are often sufficient because the privatized motion is not experienced by users in real time [27]. Physical and stream-level defenses, however, directly affect the user's interaction with the XR system [29, 60, 156]. Although task, spatial, and temporal utility provide valuable estimates of real-time performance, they cannot fully capture user comfort, perceived quality, or usability. Among the 30 physical and stream defense papers, only 10 include a user study. This represents an important gap in the literature, as many defenses intended for real-time deployment are evaluated without directly measuring how users experience them.

**(RQ10) What is the state of Open Science?** Among the 50 defense papers, only 18 provide a publicly available artifact, such as datasets, privacy-mechanism implementations, or attack code. Open science is particularly challenging in XR motion privacy because motion traces often contain biometric information, making public dataset release inappropriate or impossible in some cases [54]. However, reproducibility does not depend solely on dataset availability. Even when datasets cannot be released, authors may still be able to share privacy-mechanism implementations, attack and evaluation code, model configurations or trained models that enable others to inspect, reproduce, and extend the proposed methods. The limited availability of such artifacts makes it difficult to reproduce published results, compare defenses under common evaluation settings, and build upon prior work. Thus, while privacy concerns may justify withholding sensitive datasets, the low overall rate of artifact release remains an important reproducibility gap.

**(RQ11) What are the current research gaps in the attack and defense literature?** Across the preceding research questions, we identify the following research gaps:

- (G1) Limited research on facial and full-body motion in both the attack and defense literature (RQ1, RQ5).
- (G2) Limited research on privacy risks beyond recognition, particularly demographics, health & physiology, cognitive & affective states, and behavior, intent & preference (RQ2, RQ5).
- (G3) Limited evaluations that jointly consider multiple motion modalities and multiple privacy risks (RQ3, RQ4, RQ5).
- (G4) Limited research on physical privacy defenses compared with stream-level and aggregate-level defenses (RQ6).
- (G5) Limited evaluations of defense performance across different candidate population sizes (RQ7).
- (G6) Limited evaluations that report results under both oblivious and adaptive adversaries (RQ8).
- (G7) Limited use of user studies for physical and stream defenses that may directly affect users during interaction (RQ9).
- (G8) Limited availability of publicly released research artifacts to support reproducibility, comparison, and future extensions (RQ10).

Overall, these 8 gaps show that XR motion privacy research has made substantial progress, but important challenges remain in expanding modality and privacy-risk coverage, improving evaluation consistency, and supporting reproducibility.

## 8 Guidelines for XR Motion Privacy Research

Based on the research gaps identified throughout this SoK, we provide several guidelines for designing and evaluating future XR

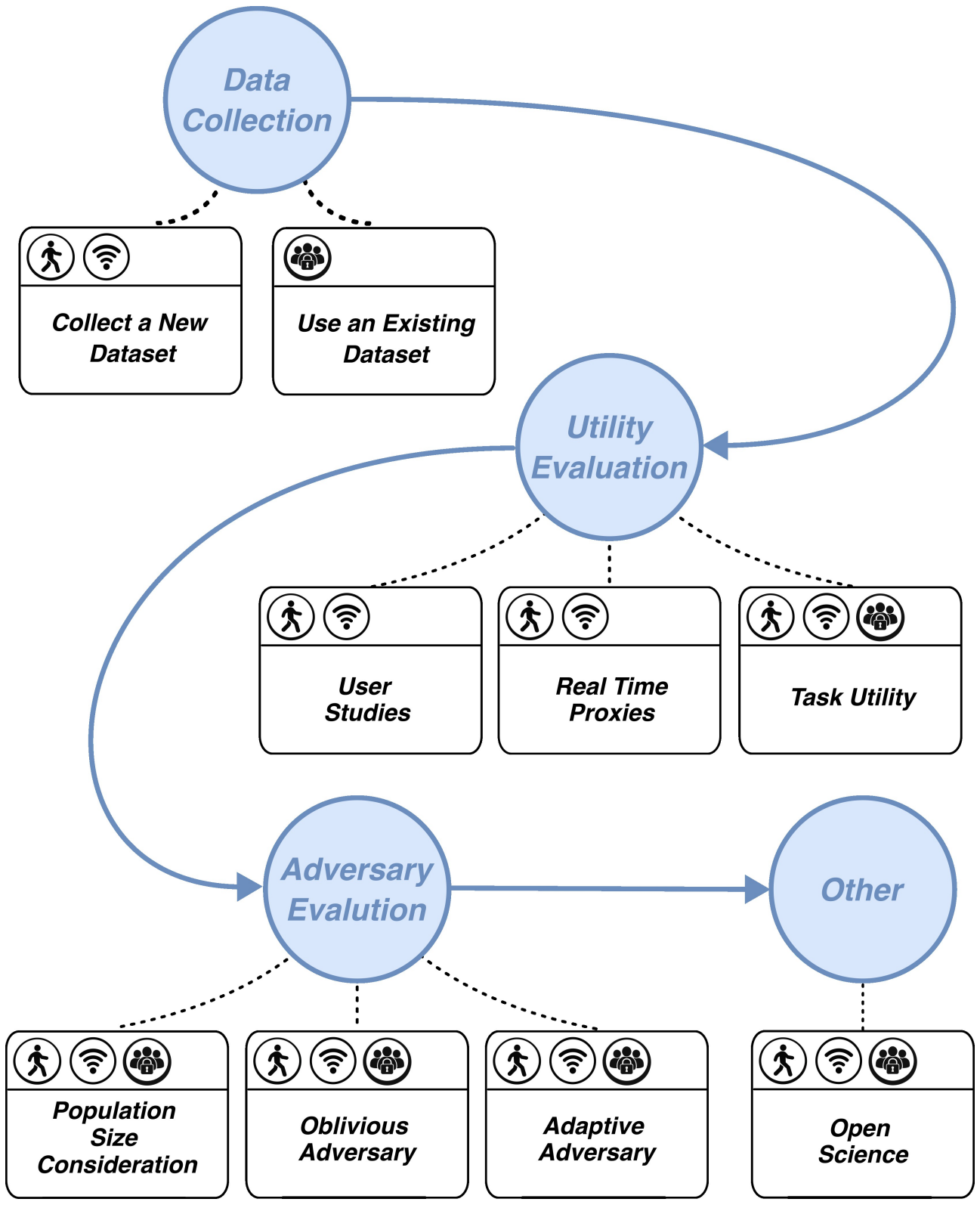

**Figure 5: Guidelines for XR privacy research. Blue circles denote guideline categories and white rectangles individual guidelines. Icons at the top of each guideline indicate the applicable defense types: physical, stream, and aggregate privacy. For example, indicates a guideline applicable to physical or stream privacy defenses.**

motion privacy research. Figure 5 summarizes these guidelines, which we describe in greater detail below.

***Data Collection.*** Data-collection requirements depend on the privacy type being studied. Our recommendations can be summarized into two main approaches: (a) collecting a dedicated dataset, and (b) using an existing public dataset.

(1) **Collecting a new dataset.** This approach is generally most suitable for physical and stream defenses. Since these mechanisms operate during user interaction (§4), collecting a dedicated dataset can often be naturally integrated into the user studies discussed next. When collecting a new dataset, we recommend recording all motion modalities available on the XR device. Even when a study focuses on a single modality, collecting additional modalities can support future multimodal privacy research without requiring another data-collection effort. This can help address the limited coverage of certain motion modalities and multimodal evaluations (G1, G3). Researchers should also consider collecting labels relevant to multiple privacy risks, such as gender, age, stress level, health status, and other information relevant to the intended privacy analyses. This can similarly support research beyond the currently dominant privacy risks and enable evaluations across multiple risks (G2, G3). We recognize that collecting additional sensitive information may complicate the IRB process; however, when appropriate and ethically justified, these labels can substantially increase the value and reusability of the collected dataset.

(2) **Using an existing public dataset.** This approach is generally most suitable for aggregate-level defenses, since these mechanisms operate on motion data that has already been recorded (§4) and therefore often do not require a new data-collection effort. When selecting a public dataset, researchers should consider the available motion modalities, the labels needed for the intended privacy risks, and the number of participants. These considerations directly reflect the gaps identified in modality and privacy-risk coverage, multimodal and multi-risk evaluation (G1, G2, G3). The selected dataset should therefore provide sufficient coverage to support the goals of the intended evaluation.

***Utility Evaluation.*** Utility evaluation also depends on the privacy type being studied. Our recommendations can be summarized into two main approaches: (a) evaluating physical and stream defenses, and (b) evaluating aggregate defenses.

(1) **Physical and stream defenses.** Since these mechanisms operate during user interaction (§4), user studies can provide substantial value by capturing how users directly experience the privacy mechanism. Given the limited use of user studies for these defense types (G7), we therefore encourage researchers to include user studies whenever practical. For examples of study designs and questions, researchers can refer to the works in Table 2 that include user evaluations. We also recommend reporting real-time utility proxies, such as spatial and temporal precision. These measures can complement user studies and help explain why users perceive or accept a privacy mechanism in a particular way. If the privacy mechanism is designed for a specific application, task utility can provide an additional useful measure. For example, a mechanism designed for Beat Saber could evaluate whether in-game scores are preserved, while a mechanism designed for foveated rendering could evaluate whether area-of-interest (AOI) accuracy is maintained.

(2) **Aggregate defenses.** Since these mechanisms do not operate during live interaction, user studies and real-time utility proxies are generally less relevant. Instead, task utility becomes particularly important for determining whether the privatized motion remains useful for its intended purpose. Examples include preserving in-game scores, AOI accuracy, or other application-specific outputs that the privacy mechanism is intended to maintain.

***Adversary Evaluations.*** At this point, the recommendations for all three privacy types converge. Our recommendations can be summarized into two main points: (a) evaluating against both oblivious and adaptive adversaries, and (b) considering the size of the candidate population.

(1) **Oblivious/adaptive adversaries.** Given the limited number of studies that report results under both adversary types (G6), we recommend evaluating against both oblivious and adaptive adversaries whenever possible. For example, in multimodal attacks, this could involve using a single motion modality to represent an oblivious adversary and multiple motion modalities to represent an adaptive adversary [6]. For other types of attacks, studies could consider two conditions: (a) privatizing the test set but not the training set, representing an oblivious adversary, and (b) privatizing both the training and test sets, representing an adaptive adversary [108].
(2) **Candidate population size considerations.** Given the limited evaluation of defense performance across different candidate population sizes (G5), whenever possible, studies should report privacy results across multiple candidate population sizes. For example, a study with 300 participants could evaluate its defense at $N = 5$, $N = 10$, $N = 20$, $N = 50$, $N = 100$, $N = 150$, $N = 200$, and $N = 300$, or at any other set of population sizes that demonstrates how the defense generalizes as the number of candidate participants varies.

***Open Science.*** Given the limited availability of publicly released research artifacts (G8), we encourage researchers to make datasets and code available whenever possible. Sharing implementation code could improve reproducibility and comparison by clarifying defense configurations, preprocessing, and evaluation procedures. Similarly, sharing datasets could reduce repeated data-collection efforts and support future evaluations of additional privacy risks, motion modalities, and defenses on the same participant data. When public dataset release is not feasible due to privacy, consent, licensing, or other ethical constraints, controlled-access models may provide an alternative. For example, researchers could grant access through data-use requests, institutional or IRB approval when necessary, and specified data-use and security conditions. When appropriate, such models could provide many of the benefits of open science while offering additional safeguards for sensitive participant data.

## 9 Opportunities for Future Work

Our systematization shows that XR motion privacy spans a broad range of motion modalities, privacy risks, adversaries, and defense approaches, while identifying several areas that remain underexplored or insufficiently evaluated. Based on these findings, we highlight several directions for future XR motion privacy research.

**(FW1) Reproduction Studies and Evaluation Frameworks.** As highlighted throughout this SoK, studies often use different adversary assumptions, utility evaluations, and study scales, making results difficult to compare directly. Reproduction studies could evaluate existing privacy mechanisms under common settings, including both oblivious and adaptive adversaries (G6), multiple candidate population sizes (G5), and user studies (G7). Evaluation frameworks could support these efforts by providing common attacks, defenses, datasets, adversary settings, and evaluation metrics, while improving artifact availability (G8). Similar frameworks exist in related biometric research [30, 34, 49], including OpenGait for gait recognition [34]. Together, these efforts would help determine whether reported privacy gains hold across evaluation conditions and improve the comparability and reproducibility of XR motion privacy research.

**(FW2) Broader Multimodal and Multi-Risk Evaluations.** As highlighted throughout this SoK, the field would benefit from evaluations that combine three properties that are currently rarely studied together: (a) large participant populations (G5), (b) all five motion modalities considered in this work (G1), and (c) labels supporting multiple privacy risks (G2). Such evaluations would make it possible to study how privacy attacks generalize across modalities and inference targets, while also providing a stronger basis for evaluating defenses under realistic multimodal exposure.

**(FW3) Physical Defenses.** Physical defenses deserve substantially more attention given their limited representation compared with stream-level and aggregate-level defenses (G4) because they are the only defense type that can protect motion before it reaches the XR device. This makes them particularly relevant to threat models in which the headset, operating system, or platform itself cannot be trusted. Future work should therefore explore a wider range of physical interventions and evaluate them across different motion modalities (G1) and privacy risks (G2).

## 10 Conclusion

This SoK presented a motion-centered systematization of XR privacy research, covering 134 publications on the privacy risks and defenses associated with motion data. We developed an XR motion threat model (§4) and taxonomy (§5) that characterize how motion is exposed to adversaries, the modalities and representations involved, the privacy risks that can be inferred, and where defenses can intervene. Our analysis shows that XR motion privacy has developed into a broad research area, but important gaps remain (§6, §7). These gaps include limited research on facial and full-body motion and privacy risks beyond recognition; limited evaluations across multiple modalities and risks; limited research on physical privacy defenses; limited evaluations across candidate population sizes and adversary assumptions; and limited use of user studies and publicly released research artifacts. Based on these gaps, we provide guidelines for data collection, utility and adversary evaluation, and open science (§8), and identify opportunities for future research (§9). Together, this systematization provides a common foundation for evaluating existing work and developing more comprehensive and reproducible XR motion privacy research.

## Ethical Considerations

This work is a systematization of prior literature. Because the work analyzes published papers, we did not seek new IRB review. We did not recruit participants, collect new human-subject data, deploy a system, or observe users in live XR environments. This review also did not include secondary analysis of existing datasets of human data.

## Open Science

To support reproducibility, we provide the screened-paper list, exclusion decisions, final corpus, and coding sheet. The materials are available at the following anonymous repository: https://anonymous.4open.science/r/SoK-Motion-Data-Privacy-in-Extended-Reality/

Additionally, recognizing the rapid growth of XR motion privacy research, we have established a webpage to support the continued systematization of future work. The webpage provides an interface through which researchers can submit newly published papers for consideration. Submitted papers are reviewed using the same inclusion criteria as this SoK and, if eligible, are added to the online corpus alongside the papers identified in our review. This living corpus is intended to keep the systematization current as the field evolves, facilitate future extensions of this SoK, and increase the visibility of relevant work within the research community. Researchers can also subscribe to updates when new papers are added, providing a way to stay informed about newly identified XR motion privacy research. The webpage link will be made publicly available upon acceptance of this paper.

## AI Use

The authors used generative AI-based tools to revise the manuscript text for clarity, flow, grammar, spelling, and typographical errors. The use of these tools was limited to manuscript text and did not extend to the bibliography. All AI-assisted edits were manually reviewed and verified by the authors. The authors take full responsibility for the accuracy, originality, and integrity of the final paper.

Image-based generative AI was also used to create some visual elements in the teaser and taxonomy figures, specifically the illustrated characters and their backgrounds. These elements were generated to avoid using photographs of real people, which could raise privacy concerns or compromise anonymous review if an image were linkable to the authors or other identifiable individuals. The generated elements are used only as illustrative visuals; all text, labels, structure, and substantive content in the figure were created and verified manually by the authors.